\documentclass[onecolumn,showpacs,amsmath,amssymb,eqsecnum]{revtex4}
\usepackage{dcolumn}
\usepackage{bm}
\usepackage[dvips]{graphicx}
\usepackage{wrapfig}
\usepackage{psfrag}
\begin{document}
\Large

\newcommand\x{{\parfillskip0pt\par}}
\newcommand\xe{{\parfillskip0pt\par}\vskip\abovedisplayskip\noindent}

\newcommand\lb{{\linebreak}}
\newcommand\noi{{\noindent}}
\newcommand\pb{{\pagebreak}}

\newcommand{\Z}{\mathbb{Z}}
\newcommand{\R}{\mathbb{R}}
\def\mK{\mathop{{\mathfrak {K}}}\nolimits}
\def\mR{\mathop{{\mathfrak {R}}}\nolimits}
\def\mv{\mathop{{\mathfrak {v}}}\nolimits}
\def\mS{\mathop{{\mathfrak {S}}}\nolimits}
\def\mU{\mathop{{\mathfrak {U}}}\nolimits}
\def\mI{\mathop{{\mathfrak {I}}}\nolimits}
\def\mA{\mathop{{\mathfrak {A}}}\nolimits}
\def\ma{\mathop{{\mathfrak {a}}}\nolimits}
\def\mH{\mathop{{\mathfrak {H}}}\nolimits}
\def\diag{\mathop{\rm diag}\nolimits}
\newcommand{\ccm}{\cal M}
\newcommand{\cE}{\cal E}
\newcommand{\cV}{\cal V}
\newcommand{\cI}{{\cal I}}
\newcommand{\cR}{{\cal R}}
\newcommand{\cK}{{\cal K}}
\newcommand{\cA}{{\cal A}}
\newcommand{\cU}{{\cal U}}
\newcommand{\cD}{{\cal D}}

\newcommand{\oV}{\overline{V}}
\newcommand{\os}{\overline{s}}
\newcommand{\opsi}{\overline{\psi}}
\newcommand{\ov}{\overline{v}}
\newcommand{\oW}{\overline{W}}
\newcommand{\oPhi}{\overline{\Phi}}
\newcommand{\bx}{\bf{x}}
\newcommand{\by}{\bf{y}}

\def\st{\mathop{\rm st}\nolimits}
\def\tr{\mathop{\rm tr}\nolimits}
\def\sign{\mathop{\rm sign}\nolimits}
\def\d{\mathop{\rm d}\nolimits}
\def\const{\mathop{\rm const}\nolimits}
\def\O{\mathop{\rm O}\nolimits}
\def\Spin{\mathop{\rm Spin}\nolimits}
\def\exp{\mathop{\rm exp}\nolimits}
\def\sh{\mathop{\rm sh}\nolimits}
\def\ch{\mathop{\rm ch}\nolimits}
\def\th{\mathop{\rm th}\nolimits}
\def\sin{\mathop{\rm sin}\nolimits}
\def\cos{\mathop{\rm cos}\nolimits}
\def\var{\mathop{\rm var}}
\def\Re{\mathop{\rm Re}\nolimits}
\def\Sp{\mathop{\rm Sp}\nolimits}
\def\kp{\mathop{\text{\ae}}\nolimits}
\def\bk{{\bf {k}}}
\def\bp{{\bf {p}}}
\def\bq{{\bf {q}}}
\def\lra{\mathop{\longrightarrow}}
\def\Const{\mathop{\rm Const}\nolimits}

\title{500-th solution of 2D Ising
model}

\author {S.N. Vergeles\vspace*{4mm}\footnote{{e-mail:vergeles@itp.ac.ru}}}

\affiliation{{Landau Institute for Theoretical Physics, Russian
Academy of Sciences,}\linebreak Chernogolovka, Moskow region,
142432 Russia }

\begin{abstract}
One more solution of 2D Ising model is found.
\end{abstract}


\maketitle

\section{Formulation of the problem}
\setcounter{equation}{0}

Let's consider 2D Ising model on the simplest regular lattice with
the partition function
\begin{eqnarray}
Z=\sum_{\{\sigma=\pm
1\}}\exp\left\{\theta\sum_{m=1}^{M-1}\sum_{n=1}^{N-1}
\left[\sigma_{m,n}(\sigma_{m+1,n}+\sigma_{m,n+1}) \right]\right\},
\nonumber \\[8pt]
m=1,\,\ldots,\,M, \quad n=1,\,\ldots,\,N.
\label{Z}
\end{eqnarray}
A pair of numbers $(m,\,n)$ enumerate the vertex of the lattice
which is at the intersection of $m$-th column and $n$-th line of
the lattice. First the partition function (\ref{Z}) was calculated
by L. Onsager \cite{1}, and then by several another authors and
methods (see, for example, \cite{2}-\cite{4}, and others
references can be found in \cite{5}). Taking into account that
$\sigma_{m,n}^2=1$, rewrite the partition function as
\begin{eqnarray}
Z=\sum_{\{\sigma=\pm 1\}}\prod_{m=1}^{M-1}\prod_{n=1}^{N-1}
\Big[\left(\ch\theta+\sh\theta\cdot\sigma_{m,n}\sigma_{m+1,n}\right)
\left(\ch\theta+\sh\theta\cdot\sigma_{m,n}\sigma_{m,n+1}\right)\Big].
\label{Z1}
\end{eqnarray}
The right-hand side of the last equation is a polynomial in the
variables $\{\sigma_{m,n}\}$, each variable $\sigma_{m,n}$ being
in the degree not higher than 4. As a result of summation in
(\ref{Z1}) all summands with odd degrees of each of these
variables reduce to zero. Thus one should take into account only
the summands which are proportional to $\sigma_{m,n}^0=
\sigma_{m,n}^2=\sigma_{m,n}^4=1$ for all $m$ and $n$. It follows
from (\ref{Z1}) that the sum is equal to the sum of all closed
contours (loops) on the lattice, generally with intersections and
self-intersections. Let's call the elementary parts of loops
connecting the nearest vertexes (say, $(m,\,n)$ and $(m+1,\,n)$ or
$(m,\,n)$ and $(m,\,n+1)$) by connections. Thus the length of the
loop is equal to the number of its connections. Number of
connections coming to each vertex can be equal to 0, 2 or 4. Thus
the partition function (\ref{Z1}) can be expressed as
\begin{eqnarray}
Z=(\ch\theta)^{2(M-1)(N-1)}2^{MN}\sum_{\text{loops}}(\th\theta)^{\nu},
\label{Z2}
\end{eqnarray}
where $\nu$ is the total number of connections of the loop.

Here an another method of calculation of partition function
(\ref{Z}) is proposed.

Let's consider the Clifford algebra with $MN$ generatives:
\begin{eqnarray}
\gamma_x\gamma_y+\gamma_y\gamma_x=2\delta_{xy}, \quad
x,\,y=1,\,\ldots,\,MN.
\label{AC}
\end{eqnarray}
Matrices $\{\gamma_x\}$ are Hermitian and I assume that their
dimensions to be $2^{MN/2}\times2^{MN/2}$, that is the minimal
possible one at given number of matrices \footnote{For simplicity
it is assumed that the numbers $M$ and $N$ are even.}. Algebra
(\ref{AC}) implies that the trace of any odd product of
$\gamma$-matrices is equal to zero, and
\begin{eqnarray}
\tr\gamma_x\gamma_y=2^{MN/2}\delta_{xy}, \quad
\tr\gamma_x\gamma_y\gamma_z\gamma_v=2^{MN/2}\big(\delta_{xy}\delta_{zv}-
\delta_{xz}\delta_{yv}+\delta_{xv}\delta_{yz}\big),
\label{AC1}
\end{eqnarray}
and so on. By virtue of algebra (\ref{AC}) any even product of
$\gamma$-matrices is reduced to the $\pm 1$ (versus the number of
their permutations) or to the product of two by two different
$\gamma$-matrices. According to (\ref{AC1}) the trace of the
product is equal to $\pm2^{MN/2}$ in the first case and to zero in
the last case.

The theory of $\gamma$-matrices in the spaces of large dimensions
as well as in Hilbert space can be found in \cite{6}.

Further we shall consider that the indices $x,\,y$ and so on
enumerate the nodes of the lattice and thus they are integer
vectors in the plane: ${\bf x}=(m,n)$; it is assumed that the
index $m$ increases on the right while the index $n$ increases up.
Thus, the matrix $\gamma_{{\bf x}}=\gamma_{m,n}$ is related to
each node $(m,n)$. We define also the elementary basic vectors
${\bf e}_1=(1,\,0)$ and ${\bf e}_2=(0,\,1)$, so that ${\bf
x}=m{\bf e}_1+n{\bf e}_2$.

The following statement is valid: the partition function (\ref{Z})
or (\ref{Z1}) can be rewritten also as
\begin{gather}
Z=2^{MN/2}(\ch\,2\theta)^{(M-1)(N-1)}\tr\Big\{\ldots
\nonumber \\[6pt]
\ldots\times\big[(\lambda+\mu\gamma_{{\bf x}}\gamma_{{\bf x}+{\bf
e}_2})(\lambda+\mu\gamma_{{\bf x}} \gamma_{{\bf x}+{\bf
e}_1})\big] \big[(\lambda+\mu\gamma_{{\bf x}+{\bf
e}_1}\gamma_{{\bf x}+{\bf e}_1+{\bf e}_2})
(\lambda+\mu\gamma_{{\bf x}+{\bf e}_1}\gamma_{{\bf x}+2{\bf
e}_1})\big]\times\cdots
\nonumber \\[10pt]
\cdots\times\big[(\lambda+\mu\gamma_{{\bf x}+{\bf
e}_2}\gamma_{{\bf x}+2{\bf e}_2})(\lambda+ \mu\gamma_{{\bf x}+{\bf
e}_2}\gamma_{{\bf x}+{\bf e}_1+{\bf e}_2})\big]\times
\nonumber \\[6pt]
\times\big[(\lambda+\mu\gamma_{{\bf x}+{\bf e}_1+{\bf e}_2}
\gamma_{{\bf x}+{\bf e}_1+2{\bf e}_2})(\lambda+\mu\gamma_{{\bf
x}+{\bf e}_1+{\bf e}_2} \gamma_{{\bf x}+2{\bf e}_1+{\bf e}_2})
\big]\times\cdots\Big\}.
\label{Z3}
\end{gather}
The expression in braces in (\ref{Z3}) is some polynomial in
$\gamma$-matrices. The statement is true if the traces of all
monomials of $\gamma$-matrices in (\ref{Z3}) are non-negative (the
positivity condition which is proved further) and
\begin{gather}
\lambda\equiv\cos\frac{\psi}{2}=\frac{\ch\theta}{\sqrt{\ch2\theta}},
\quad
\mu\equiv\sin\frac{\psi}{2}=\frac{\sh\theta}{\sqrt{\ch2\theta}}\longrightarrow
\sin\psi=\frac{2\th\theta}{1+(\th\theta)^2}.
\label{R}
\end{gather}
Note that the last expression (\ref{Z3}) is obtained from
(\ref{Z1}) by substitutions $\sigma_{m,n}\rightarrow\gamma_{m,n}$
and $\sum_{\{\sigma=\pm 1\}}\rightarrow 2^{MN/2}\tr$. But in
contrast to (\ref{Z1}) the sequence of multipliers arrangement as
well as their form are crucial in the case (\ref{Z3}). Draw
attention to the fact that in the product under the sign of trace
in (\ref{Z3}) at first the brackets are multiplied in series along
the lines, and then the results of multiplications along the lines
are multiplied in series along the column.

To establish the coincidence of the right-hand sides of Eqs.
(\ref{Z3}) and (\ref{Z2}), it is sufficient to prove the
positivity condition and to take into account the fact that under
the sign of trace in (\ref{Z3}) only that polynomials in
$\gamma$-matrixes "survive" which contain each of $\gamma$-matrix
in degrees 0, 2 or 4 (as well as in the case of variables
$\sigma$) and use designations (\ref{R}).

Let's prove positivity condition with the help of mathematical
induction method. During the process of calculations the numerical
factors are ignored since only the question about the sign of the
loop is significant here.

We begin the calculation with verification of the fact that the
matrix factor of elementary cell, corresponding to elementary loop
binding elementary cell with vertexes
\begin{gather}
{\bf x}, \quad ({\bf x}+{\bf e}_1), \quad ({\bf x}+{\bf e_1}+{\bf
e}_2), \quad ({\bf x}+{\bf e}_2),
\label{ind1}
\end{gather}
is equal to unit. Indeed, according to (\ref{Z3}) this matrix
factor is equal to
\begin{gather}
\big(\gamma_{{\bf x}}\gamma_{{\bf x}+{\bf
e}_2}\big)\big(\gamma_{{\bf x}}\gamma_{{\bf x}+{\bf e}_1}\big)
\big(\gamma_{{\bf x}+{\bf e}_1}\gamma_{{\bf x}+{\bf e}_1+{\bf
e}_2}\big) \big(\gamma_{{\bf x}+{\bf e}_2}\gamma_{{\bf x}+{\bf
e}_1+{\bf e}_2}\big)=1.
\label{ind2}
\end{gather}
The further calculation is based on the possibility of
presentation of loop matrix factor as a product of matrix factor
of smaller loop, binding smaller number of elementary cells as
compared with initial loop, and the matrix factors of elementary
cells supplying smaller loop up to the initial one. Each of the
equalities in Figs. $1\,a$ and $1\,b$ represents graphically
equality between loop matrix factor in the left-hand side and
smaller loop matrix factor times elementary cells matrix factors.
All loop matrix factors are calculated according to order in
(\ref{Z3}). In Fig. 1 solid lines describe loops while dotted
lines divide loop interiors into elementary cells. Equations in
Fig. 1 are established easily by direct testing. For example, for
Figs. $1\,a\,1$ and $2$ we have:

\psfrag{1s}{\kern-5pt\lower-1pt\hbox{\large {\sl 1}}}
\psfrag{2s}{\kern0pt\lower0pt\hbox{\large {\sl 2}}}
\psfrag{3s}{\kern0pt\lower0pt\hbox{\large {\sl 3}}}
\psfrag{4s}{\kern0pt\lower0pt\hbox{\large {\sl 4}}}
\psfrag{5s}{\kern0pt\lower0pt\hbox{\large {\sl 5}}}
\psfrag{6s}{\kern0pt\lower0pt\hbox{\large {\sl 6}}}
\psfrag{7s}{\kern0pt\lower0pt\hbox{\large {\sl 7}}}
\psfrag{8s}{\kern0pt\lower0pt\hbox{\large {\sl 8}}}
\psfrag{9s}{\kern0pt\lower0pt\hbox{\large {\sl 9}}}
\psfrag{10s}{\kern0pt\lower0pt\hbox{\large {\sl 10}}}
\begin{center}
\includegraphics[scale=1.0]{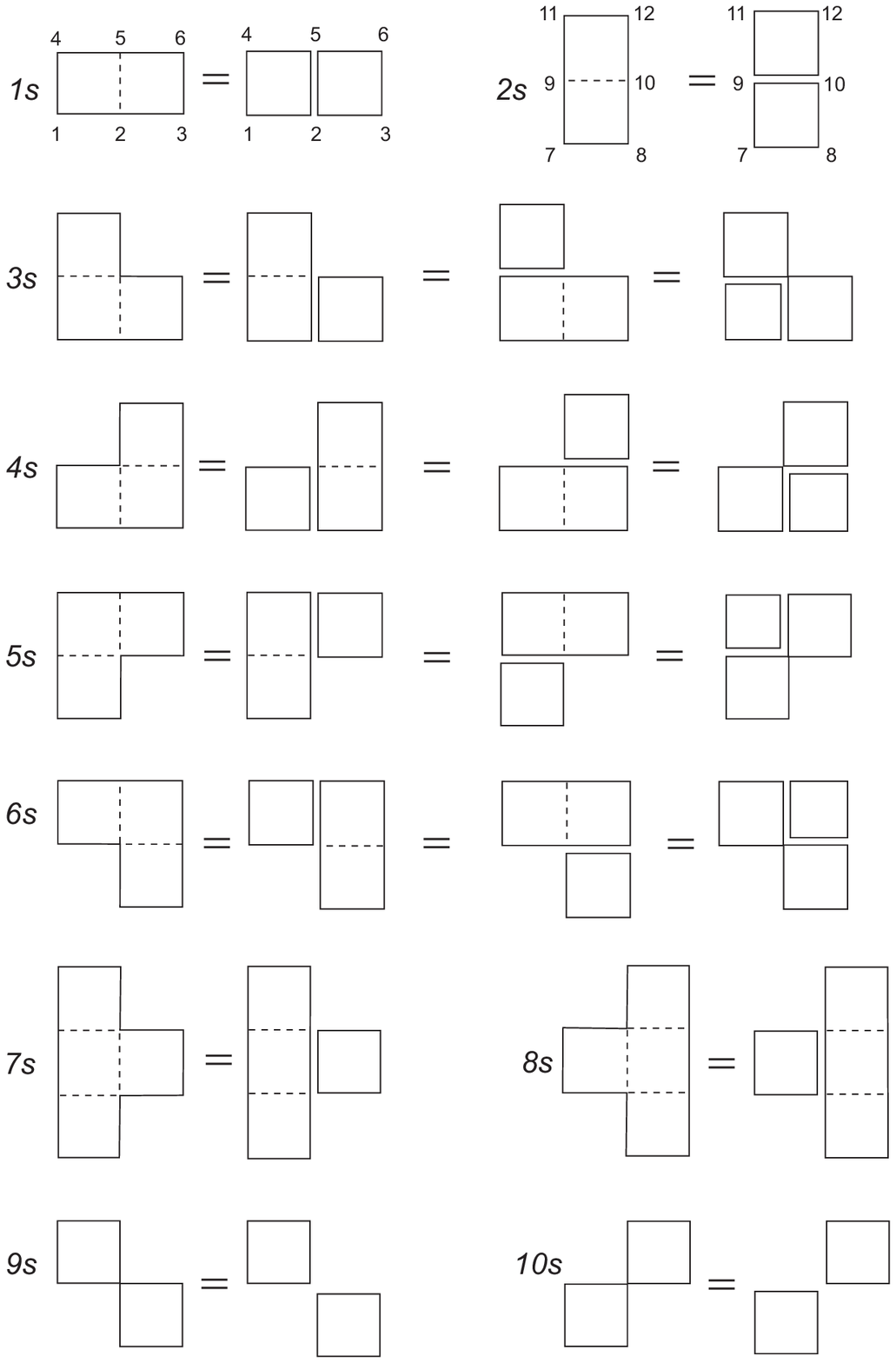}
\end{center}
\vskip8pt

\begin{center}
Fig. 1\,{\it a}
\end{center}
\vskip8pt

\psfrag{1cs}{\kern0pt\lower0pt\hbox{\large {\sl 11}}}
\psfrag{12s}{\kern0pt\lower0pt\hbox{\large {\sl 12}}}
\psfrag{13s}{\kern0pt\lower0pt\hbox{\large {\sl 13}}}
\psfrag{14s}{\kern0pt\lower0pt\hbox{\large {\sl 14}}}
\psfrag{15s}{\kern0pt\lower0pt\hbox{\large {\sl 15}}}
\psfrag{16s}{\kern0pt\lower0pt\hbox{\large {\sl 16}}}
\psfrag{17s}{\kern0pt\lower0pt\hbox{\large {\sl 17}}}
\begin{center}
\includegraphics[scale=1]{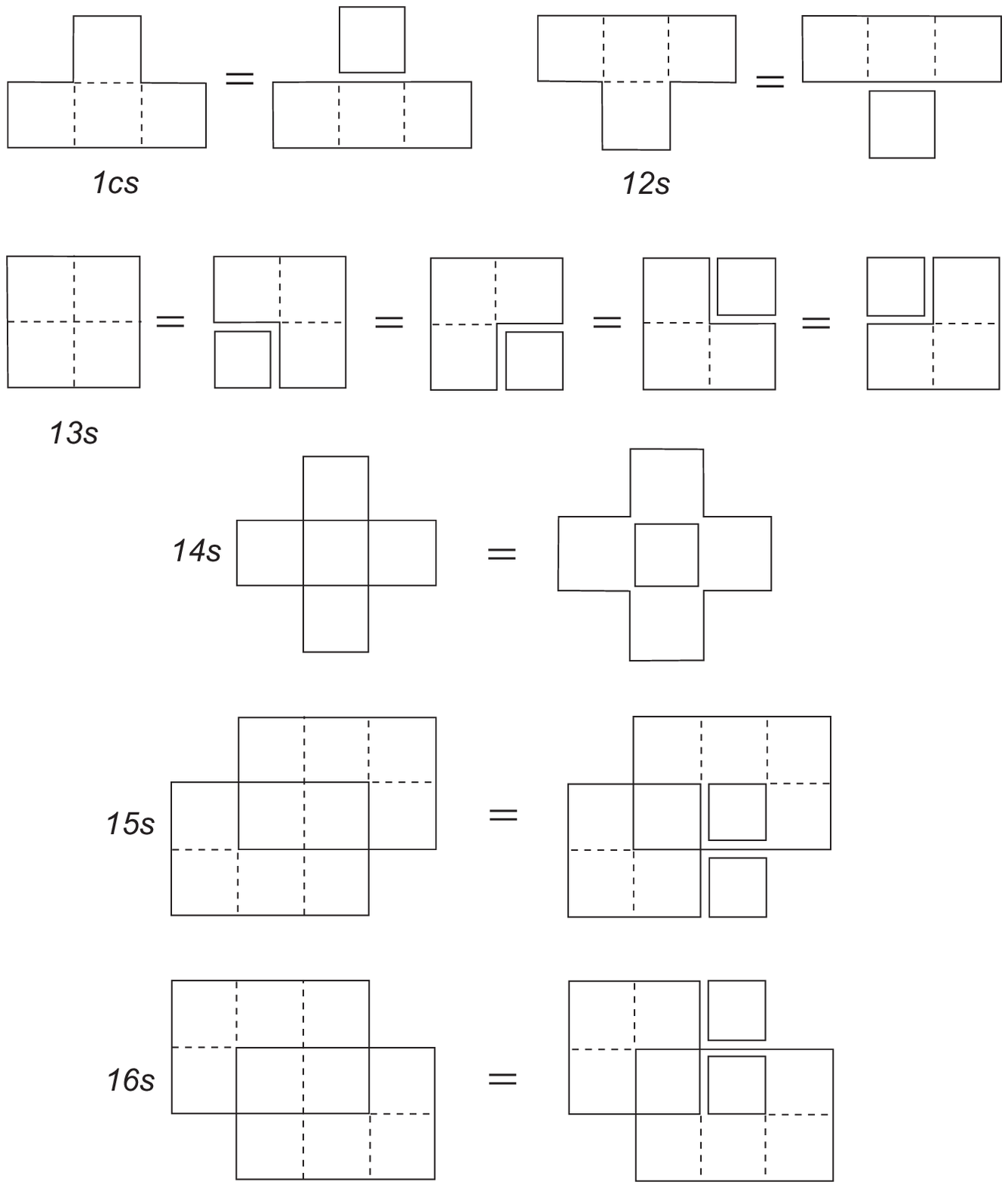}
\end{center}
\vskip8pt

\begin{center}
Fig. 1\,{\it b}
\end{center}
\vskip8pt

\begin{gather}
\big[\big(\gamma_1\gamma_4\big)\big(\gamma_1\gamma_2\big)\big(\gamma_2\gamma_3\big)
\big(\gamma_3\gamma_6\big)\big(\gamma_4\gamma_5\big)\big(\gamma_5\gamma_6\big)\big]=
\nonumber \\[8pt]
=\big[\big(\gamma_1\gamma_4\big)\big(\gamma_1\gamma_2\big)\big(\gamma_2\gamma_5\big)\big(\gamma_4\gamma_5\big)\big]
\big[\big(\gamma_2\gamma_5\big)\big(\gamma_2\gamma_3\big)\big(\gamma_3\gamma_6\big)\big(\gamma_5\gamma_6\big)\big],
\nonumber \\[8pt]
\big[\big(\gamma_7\gamma_9\big)\big(\gamma_7\gamma_8\big)\big(\gamma_8\gamma_{10}\big)
\big(\gamma_9\gamma_{11}\big)\big(\gamma_{10}\gamma_{12}\big)\big(\gamma_{11}\gamma_{12}\big)\big]=
\nonumber \\[8pt]
=\big[\big(\gamma_7\gamma_9\big)\big(\gamma_7\gamma_8\big)
\big(\gamma_8\gamma_{10}\big)\big(\gamma_9\gamma_{10}\big)\big]
\big[\big(\gamma_9\gamma_{11}\big)\big(\gamma_9\gamma_{10}\big)
\big(\gamma_{10}\gamma_{12}\big)\big(\gamma_{11}\gamma_{12}\big)\big],
\label{ind3}
\end{gather}
and so on. Developing loops and their matrix factors following to
the schemes in Fig. 1, one can realize matrix factors of any loops
as the product of elementary cells matrix factors which are equal
to unit according to (\ref{ind2}). Thus the equality of any loops
matrix factors to unit is proved inductively. Hence the positivity
condition is established.

It is evident that each multiplier $(\lambda+\mu\gamma_{{\bf
x}}\gamma_{{\bf x}+{\bf e}})$ in the braces in (\ref{Z3}) is the
matrix of orthogonal rotation by the angle $\psi$ in spinor
representation in $MN$-dimensional Euclidean space in the plane
marked by the pair of $\gamma$-matrices in the multiplier.
Therefore, all product in the braces in (\ref{Z3}) is the matrix
of an orthogonal rotation in spinor representation in
$MN$-dimensional Euclidean space.  It is known \cite{6} that this
matrix can be expressed in the form
\begin{gather}
{\cal U}=\exp\left(\frac14\,\omega_{{\bf x},{\bf y}}\gamma_{{\bf
x}}\gamma_{{\bf y}}\right), \quad \omega_{{\bf x},{\bf
y}}=-\omega_{{\bf y},{\bf x}},
\label{repr1}
\end{gather}
and
\begin{gather}
{\cal U}^{\dag}\gamma_{{\bf x}}{\cal U}={\cal O}_{{\bf x},{\bf
y}}\gamma_{{\bf y}}, \quad {\cal O}_{{\bf x},{\bf y}}\equiv
\left(e^{\omega}\right)_{{\bf x},{\bf y}}=\delta_{{\bf x},{\bf
y}}+\omega_{{\bf x},{\bf y}}+ \frac{1}{2!}\,\omega_{{\bf x},{\bf
z}}\omega_{{\bf z},{\bf y}}+\ldots. \label{repr2}
\end{gather}
The trace of the braces in (\ref{Z3}), i.e. the trace of the
matrix ${\cal U}$, is expressed simply through the eigenvalues of
real orthogonal matrix ${\cal O}_{{\bf x},{\bf y}}$. Let the set
of numbers
\begin{gather}
\left(\rho_1,\,\overline{\rho}_1,\,\rho_2,\,\overline{\rho}_2,\,\ldots,\,
\rho_{MN/2},\,\overline{\rho}_{MN/2}\right)
\label{repr3}
\end{gather}
form the complete set of eigenvalues of the matrix ${\cal O}_{{\bf
x},{\bf y}}$. Then (see Appendix)
\begin{gather}
\tr{\cal
U}=\prod_{k=1}^{MN/2}\left[2\ch\left(\frac{\ln\rho_k}{2}\right)\right]=
\prod_{k=1}^{MN/2}\left[2\cos\left(\frac{\phi_k}{2}\right)\right]=
\prod_{k=1}^{MN/2}\left(\sqrt{\rho_k}+\sqrt{\overline{\rho}_k}\right),
\nonumber \\[8pt]
\rho_k=e^{i\phi_k}.
\label{repr4}
\end{gather}

In statistical limit,  when $M,\,N\rightarrow\infty$, the problem
of matrix ${\cal O}_{{\bf x},{\bf y}}$ diagonalization simplifies
radically since the matrices ${\cal O}_{{\bf x},{\bf x}+{\bf z}}$
and $\omega_{{\bf x},{\bf x}+{\bf z}}$ become dependent only on
${\bf z}$ at relatively small distances from boundary of the
lattice. This property is known as translational invariance.
Therefore the diagonalization of these matrices is performed by
means of Fourier transformation, i.e. by passing to quasi-momentum
representation. The following complete orthonormal set of
functions on the lattice is used for that purpose:
\begin{gather}
|p\rangle\equiv\Psi_p(m)=\dfrac{1}{\sqrt{M}}\,e^{ipm}, \quad
|q\rangle\equiv\Psi_q(n)=\dfrac{1}{\sqrt{N}}\,e^{iqn},
\nonumber \\[8pt]
p=-\dfrac{\pi(M-2)}{M},\,-\dfrac{\pi(M-4)}{M},\,\ldots,\,0,\,\dfrac{2\pi}{M},\,\ldots,\,\pi,
\nonumber \\[8pt]
q=-\dfrac{\pi(N-2)}{N},\,-\dfrac{\pi(N-4)}{N},\,\ldots,\,0,\,\dfrac{2\pi}{N},\,\ldots,\,\pi,
\nonumber \\[10pt]
|{\bf k}\rangle\equiv\Psi_{{\bf k}}({\bf x})=\Psi_p(m)\Psi_q(n),
\quad {\bf k}=(p,\,q),
\nonumber \\[10pt]
\sum_{{\bf x}}\overline{\Psi}_{{\bf k}}({\bf x})\Psi_{{\bf
k}'}({\bf x})=\delta_{{\bf k}{\bf k}'} \longleftrightarrow
\sum_{{\bf k}}\Psi_{{\bf k}}({\bf x})\overline{\Psi}_{{\bf
k}}({\bf x}')= \delta_{{\bf x}{\bf x}'}.
\label{repr5}
\end{gather}

\section{Calculation of eigenvalues}
\setcounter{equation}{0}

At first let's solve a particular problem: define for ${\bf
x}=(m,\,n)$  the spinor rotation operator transforming only
$\gamma$-matrices with indexes $({\bf x}+m'{\bf e}_1$) and $({\bf
x}+m'{\bf e}_1+{\bf e}_2)$, where $m'=0,\,\pm1,\,\dots$:
\begin{gather}
{\cal U}^{(n)}\equiv \ldots\big[(\lambda+\mu\gamma_{{\bf
x}}\gamma_{{\bf x}+{\bf e}_2})(\lambda+\mu\gamma_{{\bf x}}
\gamma_{{\bf x}+{\bf e}_1})\big]\times
\nonumber \\[8pt]
\times\big[(\lambda+\mu\gamma_{{\bf x}+{\bf e}_1}\gamma_{{\bf
x}+{\bf e}_1+{\bf e}_2}) (\lambda+\mu\gamma_{{\bf x}+{\bf
e}_1}\gamma_{{\bf x}+2{\bf e}_1})\big]\ldots\,.
\label{EV1}
\end{gather}
Here we have the ordered product of spinor rotation matrixes along
only one line $n$. The corresponding orthogonal matrix is defined
just as in (\ref{repr2}):
\begin{gather}
{\cal U}^{(n)\dag}\gamma_{{\bf y}}{\cal U}^{(n)}={\cal
O}^{(n)}_{{\bf y},{\bf z}}\gamma_{{\bf z}}.
\label{EV2}
\end{gather}
According to definitions (\ref{Z3}), (\ref{repr2}) and (\ref{EV1})
\begin{gather}
{\cal U}=\ldots{\cal U}^{(n-1)}{\cal U}^{(n)}{\cal
U}^{(n+1)}\ldots, \quad {\cal O}=\ldots{\cal O}^{(n-1)}{\cal
O}^{(n)} {\cal O}^{(n+1)}\ldots\,.
\label{EV3}
\end{gather}
In the right hand sides of Eqs. (\ref{EV3}) the ordered products
of matrixes corresponding to lines take place.

Let's find the obvious expression for the matrix ${\cal
O}^{(n)}_{{\bf y},{\bf z}}$ with the help of (\ref{EV2}). This
calculation is based on the simple relations
\begin{gather}
\left(\lambda+\mu\gamma_{{\bf x}}\gamma_{{\bf x}+{\bf
e}}\right)^{\dag}\gamma_{{\bf x}} \left(\lambda+\mu\gamma_{{\bf
x}}\gamma_{{\bf x}+{\bf e}}\right)= (\cos\psi)\gamma_{{\bf
x}}+(\sin\psi)\gamma_{{\bf x}+{\bf e}},
\label{EV4}
\end{gather}
\begin{gather}
\left(\lambda+\mu\gamma_{{\bf x}}\gamma_{{\bf x}+{\bf
e}}\right)^{\dag}\gamma_{{\bf x}+{\bf e}}
\left(\lambda+\mu\gamma_{{\bf x}}\gamma_{{\bf x}+{\bf e}}\right)=
(\cos\psi)\gamma_{{\bf x}+{\bf e}}-(\sin\psi)\gamma_{{\bf x}},
\label{EV5}
\end{gather}
\begin{gather}
\left(\lambda+\mu\gamma_{{\bf x}}\gamma_{{\bf x}+{\bf
e}}\right)^{\dag}\gamma_{{\bf z}} \left(\lambda+\mu\gamma_{{\bf
x}}\gamma_{{\bf x}+{\bf e}}\right)=\gamma_{{\bf z}}, \quad {\bf
z}\neq{\bf x},\,{\bf x}+{\bf e},
\label{EV6}
\end{gather}
following directly from Eqs. (\ref{AC}) and (\ref{R}). Either
vector ${\bf e}_1$ or vector ${\bf e}_2$ can be taken instead of a
vector ${\bf e}$ in Eqs. (\ref{EV4})-(\ref{EV6}).

Consider the first case when ${\bf y}=(m,\,n)$. Then in the left
hand side of Eq. (\ref{EV2}) all spinor rotation matrixes forming
${\cal U}^{(n)\dag}$ placed to the right from
$\left(\lambda+\mu\gamma_{{\bf y}-{\bf e}_1}\gamma_{{\bf
y}}\right)^{\dag}$ and all spinor rotation matrixes forming ${\cal
U}^{(n)}$ placed to the left from the matrix
$\left(\lambda+\mu\gamma_{{\bf y}-{\bf e}_1}\gamma_{{\bf
y}}\right)$ are cancelled mutually since they do not catch on
matrix $\gamma_{{\bf y}}$. But the "facings"
\begin{gather}
\left(\lambda+\mu\gamma_{{\bf y}-{\bf e}_1}\gamma_{{\bf
y}}\right)^{\dag}\,\ldots\, \left(\lambda+\mu\gamma_{{\bf y}-{\bf
e}_1}\gamma_{{\bf y}}\right)
\nonumber
\end{gather}
transform $\gamma_{{\bf y}}$ in accordance with Eq. (\ref{EV5}) in
which one must make the identifications ${\bf e}={\bf e}_1,\,
({\bf x}+{\bf e})={\bf y}$. So we obtain the matrix element
\begin{gather}
{\cal O}^{(n)}_{{\bf y},\,{\bf y}-{\bf e}_1}=-(\sin\psi), \quad
{\bf y}=(m,\,n). \label{EV7}
\end{gather}
\vspace*{1mm}

\psfrag{f201}{\kern-4pt\lower0pt\hbox{\small {$\bf{y}+\bf{e}_2$}}}
\psfrag{f202}{\kern-12pt\lower0pt\hbox{\small
{$\bf{y}+\bf{e}_1+\bf{e}_2$}}}
\psfrag{f203}{\kern-5pt\lower0pt\hbox{\small
{$\bf{y}+2\bf{e}_1+\bf{e}_2$}}}
\psfrag{f204}{\kern-4pt\lower0pt\hbox{\small {$\bf{y}-\bf{e}_1$}}}
\psfrag{f205}{\kern0pt\lower0pt\hbox{ \small{$\bf{y}$}}}
\psfrag{f206}{\kern-4pt\lower0pt\hbox{\small {$\bf{y}+\bf{e}_1$}}}
\psfrag{f207}{\kern-4pt\lower0pt\hbox{\small
{$\bf{y}+2\bf{e}_1$}}}
\begin{center}
\includegraphics[scale=1]{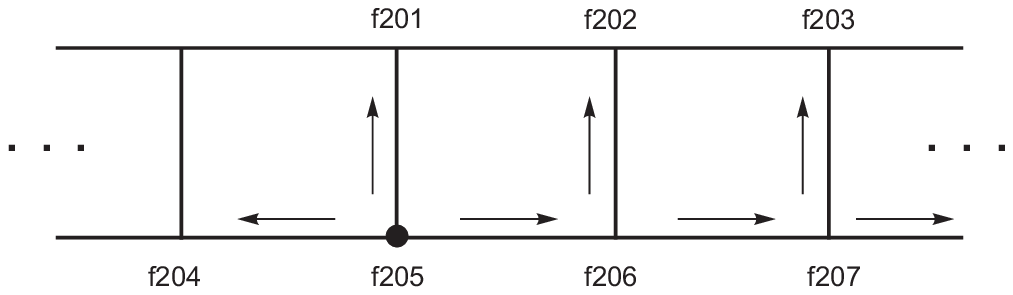}
\end{center}
\vskip1pt

\begin{center}
Fig. 2
\end{center}
\vskip3pt

Further only Eq. (\ref{EV4}) is used in which the identifications
${\bf x}=({\bf y}+m'{\bf e}_1),\,m'=0,\,1,\ldots$ are made in
series, and at each $m'$ at first ${\bf e}={\bf e}_2$ and then
${\bf e}={\bf e}_1$ are put. This process is pictured symbolically
in Fig. 2 where the bold point denotes the position of the initial
$\gamma$-matrix placed in "facings" in the left hand side of Eq.
(\ref{EV2}), and the arrows show where its image comes under the
action of the described linear transformations. Every time the
result is reduced to multiplication by $(\sin\psi)$ if the matrix
is transfered up or to the right, and to multiplication by
$(\cos\psi)$ if the matrix is left on the former place. The matrix
transfered up further remains unchanged. Thus the following result
is obtained:
\begin{gather}
{\cal O}^{(n)}_{{\bf y},\,{\bf y}+m'{\bf
e}_1}=[\sin\psi\cos\psi]^{m'}(\cos\psi)^3, \quad    {\cal
O}^{(n)}_{{\bf y},\,{\bf y}+m'{\bf e}_1+{\bf
e}_2}=[\sin\psi\cos\psi]^{m'},
\nonumber \\[10pt]
m'=0,\,1,\ldots,  \quad  {\bf y}=(m,\,n). \label{EV8}
\end{gather}

The matrix elements as ${\cal O}^{(n)}_{{\bf y}+{\bf e}_2,\,{\bf
z}}$ are calculated similarly. The process of this calculation is
pictured in Fig. 3 where the bold point again denotes the position
of $\gamma$-matrix at the beginning of the process, and the arrows
show the subsequent movements of the images of initial matrix. At
the first step the relation (\ref{EV5}) is used in which the
identifications ${\bf e}={\bf e}_2,\,{\bf x}={\bf y}$ are made.
The relation (\ref{EV4}) is used at the all subsequent steps. As a
result of simple calculation we obtain:

\begin{center}
\includegraphics[scale=1]{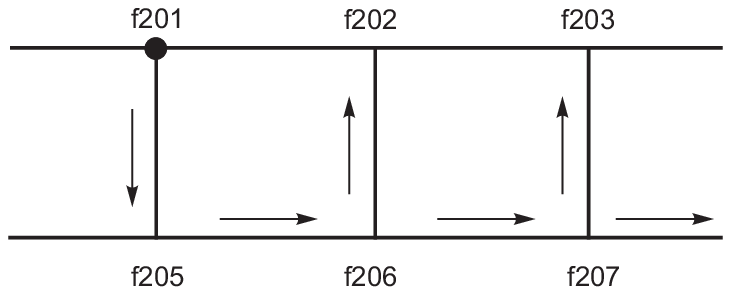}
\end{center}
\vskip1pt

\begin{center}
Fig. 3
\end{center}
\vskip3pt

\begin{gather}
{\cal O}^{(n)}_{{\bf y}+{\bf e}_2,\,{\bf y}+{\bf e}_2}=(\cos\psi),
\quad {\cal O}^{(n)}_{{\bf y}+{\bf e}_2,\,{\bf y}+(m'+1){\bf
e}_1+{\bf e}_2}=-(\sin\psi)^3 [\sin\psi\cos\psi]^{m'},
\nonumber \\[10pt]
{\cal O}^{(n)}_{{\bf y}+{\bf e}_2,\,{\bf y}+m'{\bf
e}_1}=-[\sin\psi\cos\psi]^{m'+1}, \quad  m'=0,\,1,\ldots,  \quad
{\bf y}=(m,\,n). \label{EV9}
\end{gather}

Now it is necessary to make partial diagonalization of the
orthogonal matrix using its translational invariance.  For that
purpose we make the partial Fourier transformation along the lines
of $\gamma$-matrixes with the help of Eqs. (\ref{repr5}):
\begin{gather}
\gamma_n(p)=\sum_m\overline{\Psi}_p(m)\gamma_{m,\,n}=\gamma^{\dag}_n(-p),
\quad
\left[\gamma_n(p),\,\gamma^{\dag}_{n'}(p')\right]_+=2\delta_{n,\,n'}\delta_{p,\,p'}.
\label{EV10}
\end{gather}
Let's pass to Fourier transformation along the lines in Eq.
(\ref{EV2}). Using translational invariance we rewrite this
equation in the form
\begin{gather}
{\cal U}^{(n)\dag}\gamma_{n'}(p){\cal U}^{(n)}=
\left[\sum_{m'}{\cal
O}^{(n)}_{m,n';\,m+m',n''}\Psi_p(m')\right]\gamma_{n''}(p).
\label{EV11}
\end{gather}
The matrix in square bracket in the right hand side of Eq.
(\ref{EV11}) is designated as ${\cal O}^{(n)}_{p;\,\,n',n''}$ and
it is calculated with the help of Eqs. (\ref{EV7})-(\ref{EV9}):
\begin{eqnarray}
{\cal O}^{(n)}_{p;\,\,n',n''}=
 \left(
\begin{array}{lcccccr}
\cdot & \cdot & \cdot & \cdot & \cdot & \cdot & \cdot \\
\cdot & 1 & 0 & 0 & 0 & 0 & \cdot \\
\cdot & 0 & 1 & 0 & 0 & 0 & \cdot \\
\cdot & 0 & 0 & a_p & b_p & 0 & \cdot \\
\cdot & 0 & 0 & -b_p & c_p & 0 & \cdot \\
\cdot & 0 & 0 & 0 & 0 & 1 & \cdot \\
\cdot & \cdot & \cdot & \cdot & \cdot & \cdot & \cdot
\end{array} \right)
\begin{array}{l}
 \\
 \\
 \\
\scriptstyle{n^\prime = n} \\
\scriptstyle{n^\prime = n+1} \\
\\
\\
\end{array}
. \label{EV12}
\end{eqnarray}
Here
\begin{eqnarray}
 \left\{
\begin{array}{c}
a_p \\
c_p
\end{array} \right\}  =\frac{\cos\psi-e^{\mp ip}(\sin\psi)}{1-e^{ip}[\sin\psi\cos\psi]}, \quad
b_p=\frac{[\sin\psi\cos\psi]}{1-e^{ip}[\sin\psi\cos\psi]},
\label{EV13}
\end{eqnarray}
and diagonal elements $a_p$ and $c_p$ are settled on $n$-th and
$(n+1)$-th places, correspondingly. The direct check shows that
the matrix (\ref{EV12}) is unitary.

Thus, according to (\ref{EV3})
\begin{eqnarray}
{\cal O}_{p;\,\,n',n''}\equiv\left[\sum_{m'}{\cal
O}_{m,n';\,m+m',n''}\Psi_p(m')\right]=\Big\{\ldots {\cal
O}^{(n-1)}_p{\cal O}^{(n)}_p{\cal
O}^{(n+1)}_p\ldots\Big\}_{n',n''}. \label{EV14}
\end{eqnarray}
The product (\ref{EV14}) of matrix of the kind (\ref{EV12}) is
calculated easily:
\begin{eqnarray}
{\cal O}_{p;\,\,n,n+n'}=
 \left\{
\begin{array}{cl}
0, &  \mbox{for} \ \ n'<-1 \\[5pt]
(-b_p),  & \mbox{for} \ \  n'=-1 \\[5pt]
c_pb_p^{n'}a_p,  & \mbox{for} \ \  n'\geq 0
\end{array} \right. .
\label{EV15}
\end{eqnarray}
To find the eigenvalues of orthogonal matrix (\ref{repr2}) one
must calculate the Fourier components of (\ref{EV15}) along the
column:
\begin{eqnarray}
\rho_{p,q}={\cal O}_{p,q}=\sum_{n'}{\cal
O}_{p;\,\,n,n+n'}\Psi_q(n')=
\frac{a_pc_p+b_p^2-e^{-iq}b_p}{1-e^{iq}b_p}=\frac{\eta_{p,q}}{\overline{\eta}_{p,q}},
\nonumber \\[8pt]
\eta_{p,q}=1-\left(e^{-ip}+e^{-iq}\right)(\sin\psi\cos\psi).
\label{EV16}
\end{eqnarray}
Equation (\ref{EV16}) is obtained with the help of (\ref{EV13}).

\section{The partition function}
\setcounter{equation}{0}

It follows from Eqs. (\ref{repr4}), (\ref{repr5}) and (\ref{EV16})
that free energy is proportional to the following integral
\begin{gather}
\ln\left\{
\prod_{p,q}\left(\sqrt{\rho_{p,q}}+\sqrt{\overline{\rho}_{p,q}}\right)\right\}
=\frac{MN}{4\pi^2}\int_{-\pi}^{+\pi}\d p\int_{-\pi}^{+\pi}\d q\ln
\left(\sqrt{\rho_{p,q}}+\sqrt{\overline{\rho}_{p,q}}\right)=
\nonumber \\[8pt]
=\frac{MN}{4\pi^2}\int_{-\pi}^{+\pi}\d p\int_{-\pi}^{+\pi}\d
q\ln\left(\frac{\eta_{p,q}+\overline{\eta}_{p,q}}
{\sqrt{\eta_{p,q}\overline{\eta}_{p,q}}}\right)=\frac{MN}{4\pi^2}
\int_{-\pi}^{+\pi}\d p\int_{-\pi}^{+\pi}\d
q\ln\left(\eta_{p,q}+\overline{\eta}_{p,q}\right).
\label{tt10}
\end{gather}
The last equality in (\ref{tt10}) is true because
\begin{gather}
\int_{-\pi}^{+\pi}\d p\int_{-\pi}^{+\pi}\d q\ln\eta_{p,q}=
\int_{-\pi}^{+\pi}\d p\int_{-\pi}^{+\pi}\d
q\ln\overline{\eta}_{p,q}=0.
\label{tt20}
\end{gather}
Prove the equalities (\ref{tt20}). According to (\ref{EV16})
\begin{gather}
\eta_{p,q}=\left(1-\alpha
e^{-ip}\right)\left(1-\beta_pe^{-iq}\right),
\nonumber \\[8pt]
\alpha=\frac12(\sin2\psi), \quad
\beta_p=\frac{\sin2\psi}{2-(\sin2\psi)e^{-ip}}, \quad
|\alpha|\le\frac12, \quad  \left|\beta_p\right|\le1,
\label{tt30}
\end{gather}
and $\beta_pe^{-iq}=1$ only in the case $(\sin2\psi)=1,
\,\,p=q=0$. Therefore
\begin{gather}
\int_{-\pi}^{+\pi}\d p\int_{-\pi}^{+\pi}\d q\ln\eta_{p,q}=
\nonumber \\[8pt]
=\int_{-\pi}^{+\pi}\d q\left\{\int_{-\pi}^{+\pi}\d
p\,\ln\left(1-\alpha e^{-ip}\right)\right\}+\int_{-\pi}^{+\pi}\d
p\left\{\int_{-\pi}^{+\pi}\d
q\,\ln\,\left(1-\beta_pe^{-iq}\right)\right\}=
\nonumber \\[8pt]
=-\sum_{n=1}^{\infty}\frac1n\left\{\alpha^n\int_{-\pi}^{+\pi}\d
q\int_{-\pi}^{+\pi}\d p\,e^{-inp}+\int_{-\pi}^{+\pi}\d
p\,\beta_p^n\int_{-\pi}^{+\pi}\d q\,e^{-inq}\right\}=0.
\label{tt40}
\end{gather}
The second equality in (\ref{tt20}) is proved similarly.

Eventually, we can write out the free energy with the help of
formulas (\ref{Z3}), (\ref{R}), (\ref{repr4}), (\ref{EV16}) and
(\ref{tt10}):
\begin{gather}
F=-T\ln Z=
\nonumber \\[8pt]
=-MNT\left\{\ln2+\frac{1}{8\pi^2}\int_{-\pi}^{+\pi}\d
p\int_{-\pi}^{+\pi}\d q\ln
\left[\left(\ch2\theta\right)^2-\left(\sh2\theta\right)(\cos
p+\cos q)\right]\right\}=
\nonumber \\[8pt]
=-MNT\Big\{\ln2-\ln(1-x^2)+
\nonumber \\[8pt]
+\frac{1}{8\pi^2}\int_{-\pi}^{+\pi}\d p\int_{-\pi}^{+\pi}\d q\ln
\left[(1+x^2)^2-2x(1-x^2)(\cos p+\cos q)\right]\Big\}, \quad
x=\th\theta.
\label{tt60}
\end{gather}
The last expression in (\ref{tt60}) coincides with the free energy
of 2D Ising model given in \cite{3}.

The temperature of phase transition is obtained directly from
(\ref{tt10}) and (\ref{EV16}).  Indeed, free energy has a
peculiarity at the phase transition point. It is seen from
(\ref{tt10}) that this takes place when
$\left(\eta_{p,q}+\overline{\eta}_{p,q}\right)\rightarrow 0$ for a
part of quasi-momenta, which occurs only if (see (\ref{EV16}))
\begin{eqnarray}
\sin\psi_c=\cos\psi_c=\frac{1}{\sqrt{2}},  \quad  p\longrightarrow
0, \quad q\longrightarrow 0.
\label{tt}
\end{eqnarray}
Thus with the help of (\ref{tt}) and (\ref{R}) we find that
critical temperature is found from the equation
\begin{eqnarray}
\th\theta_c=\frac{1-\cos\psi_c}{\sin\psi_c}=\sqrt{2}-1.
\label{tt3}
\end{eqnarray}

\begin{acknowledgments}

This work was supported by the Program for Support of Leading
Scientific Schools $\sharp$ 3472.2008.2.

\end{acknowledgments}

\appendix

\section{}

Here the formula (\ref{repr4}) is proved.

Let $\left\{v^{(k)}_{{\bf{x}}},
\,\,\overline{v^{(k)}_{{\bf{x}}}}\right\},\,\, k=1,\ldots,\,MN/2$,
be the complete orthonormal set of eigenvectors of the matrix
${\cal O}_{{\bf x},{\bf y}}$, so that the eigenvalue $\rho_k$
$\left(\overline{\rho}_k\right)$ corresponds to the eigenvector
$v^{(k)}_{{\bf{x}}}$ $\left(\overline{v^{(k)}_{{\bf{x}}}}\right)$.
Further also the designation
\begin{gather}
\left\{v^{(k)}_{{\bf{x}}},
\,\,\overline{v^{(k)}_{{\bf{x}}}}\right\}\equiv
\left\{v^a_{{\bf{x}}}\right\}, \quad a=1,\,\ldots,\,MN
\nonumber
\end{gather}
is used. We shall consider the introduced vectors as
vector-columns and the upper indices ${}^T$ and ${}^{\dag}$ denote
the transposition and Hermitian conjugation of vectors and
matrices. By definition
\begin{gather}
v^{(k)\,T}v^{(k')}\equiv\sum_{{\bf{x}}}v^{(k)}_{{\bf{x}}}v^{(k')}_{{\bf{x}}},
\quad
v^{(k)\,\dag}v^{(k')}\equiv\sum_{{\bf{x}}}\overline{v^{(k)}_{{\bf{x}}}}v^{(k')}_{{\bf{x}}}.
\label{ap1}
\end{gather}
The given definitions imply the following formulas:
\begin{gather}
v^{(k)\,T}v^{(k')}=0,  \quad v^{(k)\,\dag}v^{(k')}=\delta_{k\,k'},
\label{ap2}
\end{gather}
\begin{gather}
U_{{\bf{x}}\,a}\equiv v^a_{{\bf{x}}} \,\, \mbox{or}  \,\,
U\equiv\left(v^{(1)},\,\overline{v^{(1)}},\,v^{(2)},\,\overline{v^{(2)}},\,\ldots\right),
\quad \left(U^{\dag}U\right)_{ab}=\delta_{ab},
\label{ap3}
\end{gather}
\begin{gather}
\big(U^{\dag}{\cal
{O}}U\big)_{ab}=\diag\big(\rho_1,\,\overline{\rho}_1,\,\rho_2,\,\overline{\rho}_2,\,\ldots\big)
\equiv D_{ab}.
\label{ap4}
\end{gather}
It is shown in \cite{6} that
\begin{gather}
\left(U^{\dag}\omega
U\right)_{ab}=\diag\left(\ln\rho_1,\,-\ln\rho_1,\,\ln\rho_2,\,-\ln\rho_2,\,\ldots\right)
\equiv\Delta_{ab}.
\label{ap5}
\end{gather}
Due to (\ref{ap3}) and (\ref{ap5}) we have
\begin{gather}
\frac14\,\gamma_{{\bf x}}\omega_{{\bf x},{\bf y}}\gamma_{{\bf y}}=
\frac14\left(\gamma_{{\bf x}}U_{{\bf x}\,a}\right)\Delta_{ab}
\left(U^{\dag}_{b{\bf y}}\gamma_{{\bf y}}\right).
\label{ap6}
\end{gather}
$2^{MN/2}\times2^{MN/2}$-matrixes
\begin{gather}
c^{\dag}_k\equiv\gamma_{{\bf {x}}}v^{(k)}_{{\bf{x}}}, \quad
c_k\equiv\gamma_{{\bf {x}}}\overline{v^{(k)}_{{\bf{x}}}}
\label{ap7}
\end{gather}
possess all properties of fermion creation and annihilation
operators. Indeed, in consequence of (\ref{AC}) and (\ref{ap2})
\begin{gather}
[c_k,\,c^{\dag}_{k'}]_+=\delta_{kk'}, \quad
[c_k,\,c_{k'}]_+=[c^{\dag}_k,\,c^{\dag}_{k'}]_+=0.
\label{ap8}
\end{gather}
According to the definitions (\ref{ap3}) and (\ref{ap7}) we have
\begin{gather}
\gamma_{{\bf x}}U_{{\bf x}a}=
\left(c^{\dag}_1,\,c_1,\,\ldots,\,c^{\dag}_{MN/2},\,c_{MN/2}\right).
 \label{ap9}
\end{gather}
With the help of Eqs. (\ref{ap5}), (\ref{ap8}) and (\ref{ap9}) the
quantity (\ref{ap6}) is rewritten as
\begin{gather}
\frac14\,\gamma_{{\bf x}}\omega_{{\bf x},{\bf y}}\gamma_{{\bf y}}=
\frac12\sum_{k=1}^{MN/2}\left[\ln\rho_k\left(c^{\dag}_kc_k-c_kc^{\dag}_k\right)\right]=
\sum_{k=1}^{MN/2}\left[\left(\ln\rho_k\right)c^{\dag}_kc_k-\frac12\ln\rho_k\right].
\label{ap10}
\end{gather}
Equality (\ref{repr4}) follows immediately from (\ref{ap10}) since
the calculation of the trace in terms of $\gamma$-matrixes is
equivalent to the calculation of trace in terms of the
corresponding fermionic operators (\ref{ap7}).

\end{document}